\documentstyle[psfig,11pt]{article}
\newcommand{\vsc}{\vspace{15pt}}
\newcommand{\vscone}{\vspace{10pt}}
\newcommand{\ve}{\vspace{5pt}}

\newcounter{n}
\begin{document} \mbox{}
\vscone
\begin{center}
{\large \bf LOW-TEMPERATURE FAR-INFRARED ELLIPSOMETRY OF CONVERGENT BEAM} \\
\vscone
\vscone
{\large \bf A. B. Sushkov$^{\dag}$ and E. A. Tishchenko} \\
\vscone
{\sl P.L.Kapitza Institute for Physical Problems,
Russian Academy of Science,\\
Kosygina str. 2, 117973 Moscow, Russia }
\ve \\
{\sl $^{\dag}$Institute of Spectroscopy, Russian Academy of Science,\\
142092 Troitsk, Moscow region, Russia}
\end{center} \vscone \vscone \vscone \vscone

Development of an ellipsometry to the case of a coherent far infrared
irradiation,
low temperatures and small samples is described, including a decision of
the direct and inverse problems of the convergent beam ellipsometry
for an arbitrary
wavelength, measurement technique
and a compensating orientation of cryostat windows.
Experimental results are presented: for a gold film and UBe$_{13}$
single crystal at room temperature ($\lambda$=119~$\mu m$), temperature
dependencies of the complex dielectric function of SrTiO$_{3}$
($\lambda$=119, 84 and 28~$\mu m$) and of YBa$_{2}$Cu$_{3}$O$_{7-\delta}$
ceramic ($\lambda$=119~$\mu m$).
\vsc \vscone 
\vsc \vsc

\centerline{\bf 1.\ Introduction} \vscone

Ellipsometry is  a  well  known  technique$^{1,2}$  of
measurement of the optical constants  of  different  substances.
Despite of being known from the beginning of the century,
ellipsometry has begun
to develop intensively only together with lasers and computers.
Work$^{3}$ is one of the few early attempts
to develop ellipsometry into far infrared. Recent years heavy Fermion
systems and, especially, high-T$_{C}$ superconductors has given a rise
to the new efforts in the far infrared ellipsometry$^{4,5}$. In this paper we
describe our method, ellipsometer and some experimental results.
\vsc

\centerline{\bf 2.\ Far-infrared ellipsometry on small samples} \vscone
\noindent {\bf 2.1\ Parallel or convergent beam?} \vscone

 Far infrared is characterized by significant divergency of the beams
due to the value of $\lambda$/$d$ ratio, where $\lambda$ is a wavelength
and $d$ is a beam diameter. It means, that it is impossible to create
a thin parallel beam of far infrared as it is possible in visible.
Thus, there are two alternative approaches to the far infrared
ellipsometry on the small samples:\\
1) the sample is in a wide parallel beam 
the electromagnetic field is of nonzero value on the edges of the sample,
reflection and diffraction parts of the field are detected;\\
2) the sample is placed in the focus of a lens or a mirror  
the field on the edges of the sample can be put equal to zero and
diffraction effects may be neglected. \vscone

 Both these approaches suggest appropriate calculations of diffraction
effects in the former and of the convergency of the beam in the latter case.
In the case of a wide parallel beam diffraction field can be introduced
in the parabolic-equation approximation of the scalar theory of diffraction as
parabolic-like half-shadow zones from each edge of the sample$^{6}$.
At large angles of incidence (80$^\circ$) these zones will be essentially
inter-mixed just near the far edge of the sample, so the detected field will
contain a diffraction part. Scalar theory of diffraction is not the sufficient
one for the ellipsometry. This case the decision of the problem of a polarized
plane wave diffraction on the impedance edge should be used.
We suggest that such a decision can depend on
the curvature radius of the edge as parameter, which is always unknown value
in the experiment. That item needs a separate experimental study.
Unfortunately, our sample holders ab initio were designed for
the convergent beam geometry and do not allow us to carry out such a study.\vscone

Is that diffraction effect from the edges a negligible one? To our
opinion, that effect can be neglected only in the case of the low reflected
samples. In the case of highly reflected samples the diffraction effect
of the edges
can be of the order of the small reflection effect to be measured. For example,
gold mirror$^{7,8}$ at $\lambda$=100 $\mu m$ and 80$^{\circ}$ due to reflection
produces
$\Delta_{ps}$=1.17$^{\circ}$ and $\Psi$=44.65$^{\circ}$, where as usual
$\rho=\tan\Psi\,exp(i \Delta)$.
We have chosen a convergent beam (CB) ellipsometry.
In the case of CB-ellipsometry
we reduce to a minimum value the diffraction effects, but the
problem of a correct account of the convergent beam reflection arises.
We have found
\par
\centerline{\psfig{figure=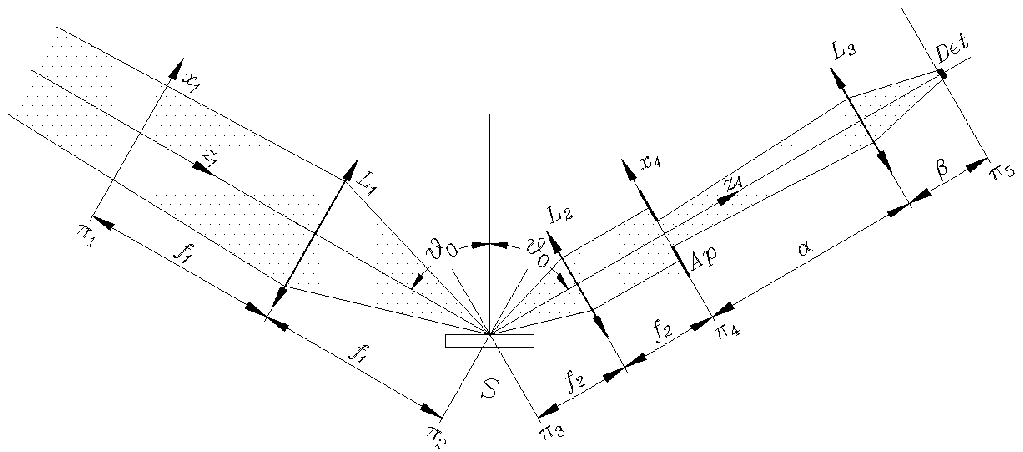}}
\begin{center}
\begin{minipage}{110mm}
\small
Fig. 1. Optical scheme of the convergent-beam ellipsometer for far infrared:
$L_{1},L_{2},L_{3}$---lenses with focal lengths $f_{1},f_{2},f_{3}$,
respectively; S---sample; Det---detector; $\pi_{1}-\pi_{5}$---
fundamental planes; $\theta_{0}$---angle of incidence of the beam; the optimal
position of polarizer and analyzer is between lenses $L_{1}$ and $L_{2}$;
and $1/\alpha+1/\beta=1/f_{3}$.
\normalsize
\end{minipage}
\end{center}
\vspace{3mm}
the decision of the direct problem of the CB ellipsometry,
quite adequate to the experimental problems.
\vscone

\noindent {\bf 2.1\ Convergent beam ellipsometry} \vscone

 The direct problem of the convergent beam ellipsometry is
to calculate the detected signal in terms of the reflection matrix of
the sample and the field distribution in the beam.
Our decision$^{4}$ is based on the lens's
Fourier-transform capability$^{9}$.
That decision simultaneously determines the optical scheme
of the CB-ellipsometer (Fig.~1).
Let's designate $E_{x_{1}}(x_{1},y_{1})$ and $E_{y_{1}}(x_{1},y_{1})$
the complex amplitudes of $x$-  and $y$-polarized beams in the input plane
$\pi_{1}$, then the field in the output plane $\pi_{4}$ is:
\begin{equation}
\left[ \begin{array}{c} E_{x_{4}} \\ E_{y_{4}} \end{array} \right]
 (x_{4},y_{4}) = {\bf R}_{C}(x_{4},y_{4})
\left[ \begin{array}{c} E_{x_{1}} \\ E_{y_{1}} \end{array} \right]
 (x_{4} f_{1}/f_{2},-y_{4} f_{1}/f_{2}) ,       \label{eq.Ex4}
\end{equation}
where ${\bf R}_{C}$ is modified reflection matrix. It can be found in coordinates
($\theta,\varphi$), where $\theta$ is the angle of incidence of a partial
plane wave in its own plane of incidence and $\varphi$ is the azimuth of this
plane
\begin{equation}
{\bf R}_{C}(\theta,\varphi)=
{\bf A}_{L}{\bf R}(\theta,\varphi)\, {\bf A}_{R}(\theta,\varphi)
\end{equation}
where
\begin{equation} {\bf R} =
\left[ \begin{array}{cc} R_{pp} & R_{ps}\\ R_{sp} & R_{ss}\end{array} \right] -
\end{equation}
is a reflection matrix of a general kind$^{1}$ for a  partial
plane  wave, and
\begin{equation} {\bf A}_{L} =
\left[ \begin{array}{rr} a & -b \\ -c &  d   \end{array} \right] , \hspace{1cm}
 {\bf A}_{R} = \frac{1}{a d - b c}
\left[ \begin{array}{rr} d & -b \\ -c &  a    \end{array} \right] ,
\end{equation}
\begin{equation} \left\{   \begin{array}{rcrl}
a&=& &\!\!\! \cos(\theta_{0}) \cos(\theta) \cos(\varphi)+
             \sin(\theta_{0}) \sin(\theta) \\
b&=&-&\!\!\! \cos(\theta_{0}) \sin(\varphi) \\
c&=& &\!\!\! \cos(\theta    ) \sin(\varphi) \\
d&=& &\!\!\! \cos(\varphi) \end{array} \right.  \label{eq.abcd}
\end{equation}
To calculate ${\bf R}_{C}$ - matrix in $(x4, y4)$ coordinates,
the variables are to be changed:
\begin{equation} \left\{  \begin{array}{l}
\cos(\theta) = \left\{ x_{4} \sin(\theta_{0})+
\cos(\theta_{0}) \sqrt{f_{2}^{2}-x_{4}^{2}-y_{4}^{2} } \right\} /f_{2} \\
\tan(\varphi)= y_{4}/ \left\{\sin(\theta_{0}) \sqrt{f_{2}^{2}-x_{4}^{2}-
y_{4}^{2} } - x_{4} \cos(\theta_{0})   \right\} \end{array} \right.
\label{ch.3}
\end{equation}

Placing an aperture in the output plane $\pi_{4}$ we can filtrate some
part of the output field, corresponding to  the  part  of  reflected
plane waves. Lens $L_{3}$ creates  the  image  of  the  aperture in  the
detector plane $\pi_{5}$. Detected signal in the ideal ellipsometer  can
be written as:
\begin{equation}
I_{D}(A) \propto  \int\!\!\!\int_{Ap}\!\{ E_{x_{4}}(x_{4},y_{4}) \cos(A) +
E_{y_{4}}(x_{4},y_{4})\sin(A)\}\{\ldots\}^{*}\,dx_{4}\,dy_{4}\label{eq.Idet}
\end{equation}
where $A$ --- angle of analyzer, $\{\ldots\}^{*}$ means complex
conjugation and the integral is to be  taken  over  the
aperture $Ap$ in the plane $\pi_{4}$. To compare the detected signal
in the case of the convergent beam to that one in the case of
the alone plane wave expression~(\ref{eq.Idet}) should be rewritten:
\begin{equation}
I_{D}(A) \propto I_{xx} \cos^{2}(A) + I_{xy} \sin(A) \cos(A) +
I_{yy} \sin^{2}(A)  \label{eq.Idet2}
\end{equation}
In the case of alone plane wave the detected signal has the same
appearance as (\ref{eq.Idet}) with  the following differences:
\newlength{\pwb} \settowidth{\pwb}{convergent beam:}
\begin{eqnarray} & \parbox[r]{\pwb}{convergent beam:} & \left\{  \begin{array}{l}
I_{xx}=\int\!\!\!\int_{Ap}\! E_{x_{4}} E^{*}_{x_{4}}dx_{4}\,dy_{4}\\
I_{xy}=\int\!\!\!\int_{Ap}\! (E_{x_{4}}  E^{*}_{y_{4}}
                     +E^{*}_{x_{4}} E_{y_{4}})\, dx_{4}\,dy_{4}\\
   I_{yy}=\int\!\!\!\int_{Ap}\!  E_{y_{4}}  E^{*}_{y_{4}} dx_{4}\,dy_{4}
         \end{array} \right. \label{eq.IxyCB} \\
& \parbox[r]{\pwb}{plane wave:} & \left\{  \begin{array}{l}
   I_{xx}= E_{x_{4}}  E^{*}_{x_{4}}  \\
   I_{xy}= E_{x_{4}}  E^{*}_{y_{4}} + E^{*}_{x_{4}} E_{y_{4}} \\
   I_{yy}= E_{y_{4}}  E^{*}_{y_{4}}  \end{array} \right. \label{eq.IxyPW}
\end{eqnarray}
If a sample is an ideal mirror, expressions (\ref{eq.IxyCB}) coincide with
(\ref{eq.IxyPW}) ones. In the opposite case, each point of the output
plane $\pi_{4}$ has its own polarization ellipse and averaged coefficients
(\ref{eq.IxyCB}) should be computed. \vscone 

In addition to the quantitative results above
two qualitative notices can be done.\\
a) The convergency of the beam will affect mainly on the averaged ellipticity
of the beam. \\
b) No matter in what point between the lenses $L_{1}$ and $L_{2}$, together
consisting a telescope, the reflecting sample is placed.
\vsc

\noindent {\bf 2.2\ Possibility of Multiple-Angle-of-Incidence (MAI) 
ellipsometry} \vscone

 Optical scheme of the CB-ellipsometer (Fig. 1) contains
an interesting opportunity of MAI measurements. To carry out MAI measurements
someone should move the output aperture $Ap$ in the plane $\pi_{4}$ in
x-direction perpendicularly to the optical axis. MAI ellipsometry can be used
to determine the thickness of the film to be investigated. We checked such
opportunity on pure substrate of SrTiO$_{3}$. Some of results are
in Fig.~2. Parameter $\chi$ follows to the model computation line, while
parameter $\gamma$ deflects from it's line appreciably. Parameter $\gamma$
is mainly connected with a phase shift. It points out to the phase distribution
in a real beam as a possible reason for that deflection. So, we can't use that
simple scheme for MAI ellipsometry, although measurements at different central
angles of incidence give more close results$^{4}$. We suggest that measurements
with uncoherent light can give better results.
\vscone \vscone

\centerline{\bf 3.\ Ellipsometric technique} \vscone
\noindent
{\bf 3.1\ Ellipsometric scheme} \vscone 

We have chosen compensatorless scheme of  the  ellipsometer  to
simplify experiment, to work in a wide spectral region,
and to avoid additional difficulties in  solving
the direct problem of CB-ellipsometry.  We  measure
two  parameters  of  the  polarization  ellipse  -  azimuth $\chi $   and
ellipticity $\gamma $ -  both averaged  over  the  aperture.
Experimentally measured values are
analyzer angle $A_{min}$, at which the detected signal is of
minimal value and relation of the minimal signal to the maximal one:
\newpage
\begin{eqnarray}   \chi & = & A_{min}                        \label{eq.chi}\\
\tan^{2}(\gamma) & = & I_{D}(A_{min})/I_{D}(A_{min}\pm\pi/2) \label{eq.gam}
\end{eqnarray}
These values can also be computed from (\ref{eq.Idet2}),~(\ref{eq.IxyCB}).
Equating  the first derivative of the detector signal (\ref{eq.Idet2}) to zero,
someone obtains:
\begin{eqnarray}
A_{ext1}=\frac{1}{2}\arctan( \frac{I_{xy}}{I_{xx}-I_{yy}}) \label{eq.ext1} \\
A_{ext2}=A_{ext1} \pm \pi/2                     \label{eq.ext2}
\end{eqnarray}
where $A_{ext1}$ and $A_{ext2}$ are extrema of the function $I_{D}(A)$, one  of
those is minimum, another --- maximum. In any case, to  use (\ref{eq.gam}) the
function (\ref{eq.Idet2}) should be computed
at both points (\ref{eq.ext1}) and (\ref{eq.ext2}).  Computer
program can readily distinguish which one is minimum and vise versa.
Thus,  we  can as measure as calculate two  parameters:
averaged azimuth $\chi $ and averaged ellipticity $\gamma $, that
permits to decide the inverse problem. \vsc \\
{\bf 3.2\ Inverse problem} \vscone 

In brief, the numerical  inversion  problem is
to find  such  values  of  parameters  of  the  reflecting
system, that error function $F$ has a minimum  value$^{1}$. For the
multiple-angle-of-incidence (MAI) technique let us  define the error function
$F$ as follows:
\begin{equation} F =\sum_{i\,=\,1}^{M}\left\{(\chi_{i}^{m} - \chi^{c}_{i})^{2}
+ (\gamma ^{m}_{i} - \gamma ^{c}_{i})^{2} \right\}
\end{equation}
where $M$ is the number of measurements (the number of incidence
angles), $\chi^{m}_{i}$  and  $\gamma^{m}_{i}$ denotes angles from
the ith measurement, $\chi^{c}_{i}$ and $\gamma^{c}_{i}$
are computed values.  In  this  paper
only a simplest  reflecting  system~---  an isotropic  semi  space~---  is
discussed. In that case  $F$ is a function of two variables, e.g.
$n$ and $\kappa$ --- refraction and extinction coefficients of the substance.
It's necessary to study the surface of $F(n,\kappa)$ to draw the
conclusion about existing of the single minimum of this function. \vscone

Let's take SrTiO$_{3}$ single crystal as probe material, because of its
large optical constants in far infrared, cubic structure and wide use as a
substrate for YBaCuO films. Results of a simple computer experiment are
in Fig.~3. Let's take the computed for the convergent beam values
$\chi^{m}$  and  $\gamma^{m}$ as 
\par
\centerline{\psfig{figure=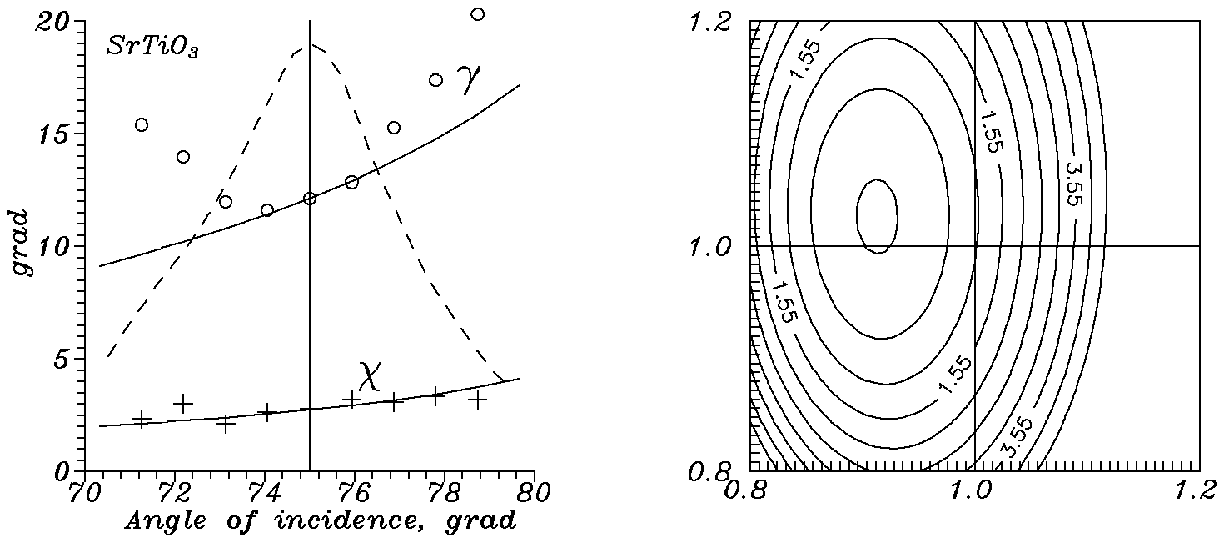}}
\small
\begin{minipage}[t]{57mm}
Fig. 2. Ellipsometric parameters $\chi$ and $\gamma$ vs a shift of
a~$6\times40$~mm
slit  in the plane $\pi_{4}$ of Fig.~1; crosses and circles ---
experiment, solid lines --- computation, dashed line --- intensity of the
beam at s-polarization.
\end{minipage} \ \hspace{9mm} \
\begin{minipage}[t]{50mm}
Fig. 3. Contour map of
the error function $F(n,\kappa)$ at $n$=22, $\kappa$=13, $\lambda$=119 $\mu m$, M=1,
$\theta_{0}$=80$^{\circ}$, aperture of diameter 30 mm, $L_{1}$=$L_{2}$=153~mm.
\end{minipage}
\vspace{3mm}
\normalsize
measured ones, but
$\chi^{c}(n,\kappa)$  and  $\gamma^{c}(n,\kappa)$ we will calculate for the
plane wave. This is a model of a convergent beam experiment with a plane wave
treatment.
\noindent Two conclusions can be drawn from Fig.~3:  

i) the error function $F(n,\kappa)$
has a single minimum, so the inverse problem has a single decision; 

ii) the displacement of the minimum illustrates the influence
of the focusing of the far infrared beam or difference between two approximations:
of the plane wave and of the convergent beam. \vscone 

Comparing $\chi=11.18^{\circ}$, $\gamma=7.02^{\circ}$, computed for the
convergent beam, to $\chi=11.19^{\circ}$, $\gamma=5.88^{\circ}$, computed
for the plane wave, we can draw the third conclusion: 

iii) the convergency of a beam influences on value of $\gamma$ and practically
does not influence on value of $\chi$.\vsc \\
{\bf 3.3\ Problem of cryostat windows} \vscone 

In a general case window is a thick anisotropic  plate. In the case of 
a convergent beam windows must be {\sl thin.} It  means  that
1-st,\, 2-nd,\, \ldots ,\, ith convergent beams, reflected inside  window  plate, all
must have the same focal point. Because of this reason, due  to  its
transparency in  visible  and  far infrared  and  also  due  to  its  low
temperature properties we use 20 $\mu m$ mylar as material both for
warm and  cold  windows.  Mylar is known to be an
anisotropic material in far infrared$^{10}$. Rotating 20 $\mu m$ mylar foil between
crossed polarizer and analyzer ($P\bot A$)
we detected two approximately perpendicular axises
of zero effect to the polarization of the beam. Maximum signal in that
experiment was $8\times10^{-4}$ of the signal in $P\|A$ orientation at
$\lambda$=119~$\mu m$. Black polyethylene shows $1.4\times10^{-3}$ at
$\lambda$=84~$\mu m$. Being mounted in an arbitrary orientation 4 windows and
2 filters can give a significant polarization effect. We proposed following
minimization and account of this effect.
\vscone \\
According to our development above, in presence of the  windows,
matrix ${\bf R}_{C}(x_{4},y_{4})$ in equation~(\ref{eq.Ex4})
should be modified:
\begin{equation}
{\bf R}_{C}(x_{4},y_{4}) \Rightarrow  T_{4}(x_{4},y_{4}) \,
T_{3}(x_{4},y_{4})\, {\bf R}_{C}(x_{4},y_{4}) \,
T_{2}(x_{4},y_{4})\, T_{1}(x_{4},y_{4})
\end{equation}
where $T_{i}(x_{4},y_{4})$ is $2\times 2$ matrix of each window
for every partial plane wave. But a strict account of the windows is
practically impossible because we deal with the not normal incidence of
a convergent beam onto a concave or convex anisotropic film
with an unknown model of anisotropy. We prefer to minimize the windows
anisotropy effect by their compensating orientation as it is described
in Appendix and
measure the rest of the effect due to the not normal incidence of the beam as
a complex constant $T_{W}$, describing the transmission of all for windows.
This value is used in solving of the inverse problem for the cold measurements.
\vsc \vscone

\centerline{\bf 4.\ Experiment} \vscone
\noindent{\bf 4.1\ Setup} \vscone 

According to the optical scheme of Fig.~1, far infrared ellipsometer was made up with
water vapor electro-discharged laser being the source and liquid helium
cooled photo-resistors Ge:Ga, Si:B, GaAS as the far infrared detectors.
The section of the setup by plane of incidence
is shown in Fig.~4c. All lenses are made up of polyethylene.
 Both polarizer and analyzer consist of two sheets:
i)\, alumina film stripes
on thephlon replica with .8 $\mu m$ period and ii)\, metal mesh with periods
20$\times$400 $\mu m$. Polarizers should be placed as close to the sample
as possible
to minimize polarization inhomogeneity of the beam due to that one of
the optical elements, including polarizer and analyzer.
The sample is
in the atmosphere of a cryoagent. To filtrate 
\par
\centerline{\psfig{figure=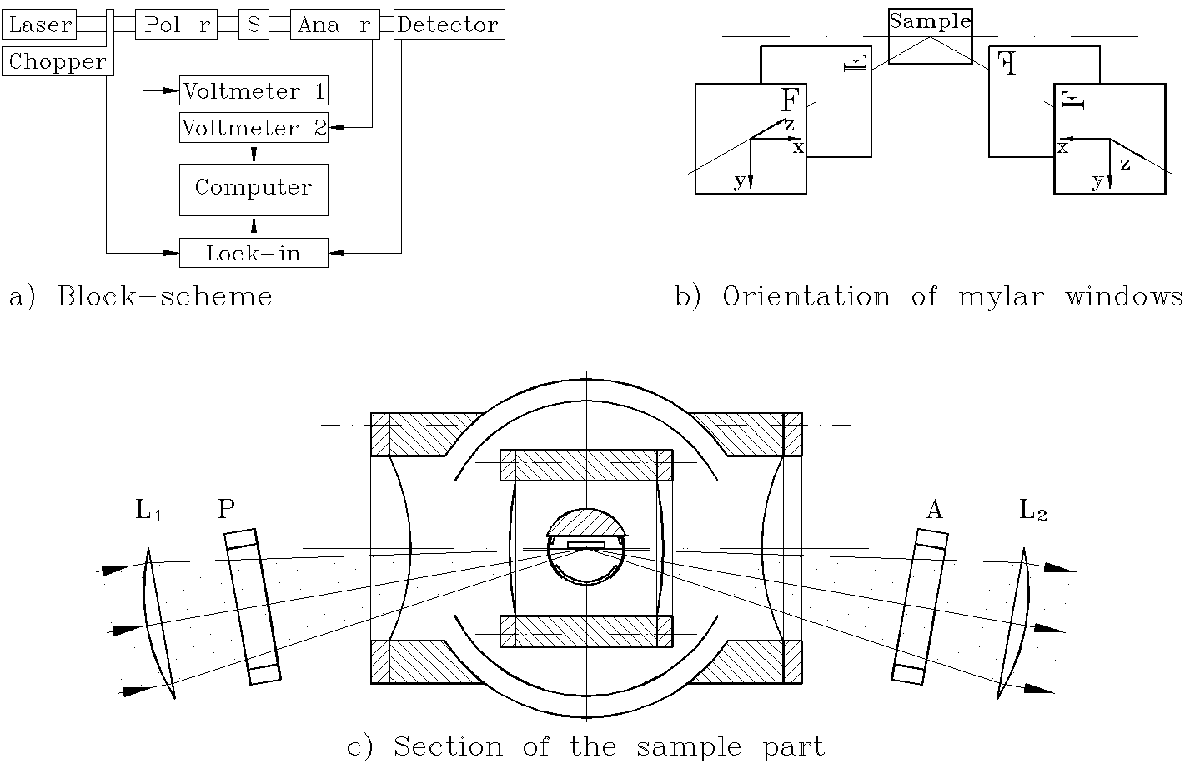}}
\begin{center}
{\small Fig. 4. Setup.}
\end{center}
\vspace{5mm}
\normalsize
the short wavelength irradiation
a black polyethylene (in compensating orientation) surrounds the
sample. \vscone

Block-scheme of the setup is in Fig.~4a.
By means of precision multi-revolution
potentiometers $P$- and $A$-angles are transformed to the voltage. Accuracy
and linearity of this transformation both are of order $\pm0.1^{\circ}$.
To measure an intensity of light
a lock-in technique at modulation frequency 80 Hz is used. The polarizer is
at fixed angle $\pm45^{\circ}$, the analyzer is rotated near the minimum of
intensity and this part of the curve $I_{D}(A)$ is recorded in the computer
file.  $A_{min}$ and $I_{D}(A_{min})$ are calculated and analyzer begins to
rotate to the $A_{max}$ angle. A linearity range of the detector does not
exceed $10^{2}$, but we need to measure values of $I_{max}/I_{min}$ of order
$10^{4}$ for the metallic samples. So we use paper sheets as
attenuators (each of order 3 at 119~$\mu m$)
to measure the signal at the same gain of
the lock-in amplifier. Such measurement routine requires a time stability of
the laser irradiation. \vscone 

The time instability of irradiation and the small movements of
the sample
in the cooling process (up to $0.3$ mm) are two main sources of random errors.
The whole
cryostat is mounted on the movable optical table to adjust the sample in
$x$-direction. Errors due to the beginning
of cooling process are good seen in the upper curves in Fig.~5. \vscone \\
\noindent{\bf 4.1.1\ Room temperature results} \vscone 

Some experimental results are presented in the Table to illustrate
the sensitivity of the ellipsometer. All these results were obtained at room
temperature, at $\lambda$=119~$\mu m$ without windows. Expected values
for gold$^{7,8}$ are
$\chi=45^{\circ}-0.36^{\circ}$ and $\tan^{2}(\gamma)=9.64\times10^{-5}$.
As can be seen
from a transmittance measurements, accuracy of the ellipsometer is sufficient
for gold's measurements.
Increasing
value of ellipticity $\gamma$, corresponding mainly to the phase shift,
can be caused by uncertainties of the surface of the gold film.
For  UBe$_{13}$ we have computed
$W_{p}=14600$~cm$^{-1}$ and $W_{\tau}=670$~cm$^{-1}$. Computed from these
values free electrons's concentration ($m_{e}=1$) $N_{e}=2.35\times10^{21}$ is
in good agreement with that one from Hall-effect measurements$^{11}$.
\begin{table}[h]
\begin{center}
\begin{tabular}{ l|l|l l l l }
\hline
Sample        & Measr.      & \multicolumn{4}{c}{Polarizer angle, P} \\
                              \cline{3-6}
              & values      & $-45^{\circ}$   &
                $0^{\circ}$ & $+45^{\circ}$   & $+90^{\circ}$  \\
\hline
\mbox{} & \mbox{} & \mbox{} & \mbox{} & \mbox{} & \mbox{} \\
No sample,    & $\chi$               & $+0.13^{\circ}$ &
                $+0.25^{\circ}$      & $-0.10^{\circ}$    & $0^{\circ}$  \\
transm.       & $\tan^{2}(\gamma)$   & $ 5 \times 10^{-6}$ &
                $ 5 \times 10^{-6}$ & $ 5 \times 10^{-6}$ &
    $ 5 \times 10^{-6}$ \\
\mbox{} & \mbox{} & \mbox{} & \mbox{} & \mbox{} & \mbox{} \\
  Au          & $\chi$               & $+0.39^{\circ}$ &
                $-0.01^{\circ}$      & $-0.60^{\circ}$    & $0^{\circ}$  \\
 film         & $\tan^{2}(\gamma)$   & $4.14 \times 10^{-4}$ &
                $9 \times 10^{-6}$ & $6.28 \times 10^{-4}$ &
    $7 \times 10^{-6}$ \\
\mbox{} & \mbox{} & \mbox{} & \mbox{} & \mbox{} & \mbox{} \\
 UBe$_{13}$   & $\chi$               & $+3.53^{\circ}$ &
                                     & $-3.41^{\circ}$    &              \\
single cr.     & $\tan^{2}(\gamma)$   & $4.74 \times 10^{-3}$ &
                                      & $4.53 \times 10^{-3}$ &       \\
\end{tabular}
\end{center}
\end{table}
\vscone \\
\noindent{\bf 4.2\ Results of SrTiO$_{3}$  measurements} \vscone 

Temperature dependencies of real and imaginary parts of the complex dielectric
function of SrTiO$_{3}$ single crystal are shown in Fig.~5. Such kind
of behavior is caused by softening of low-frequency mode,
centered between 80 -- 90 cm$^{-1}$. We have measured different samples from
independent sources and obtained different optical constants, especially at
the frequency 84 cm$^{-1}$. Re-polishing of the samples does not change their
optical constants. In literature$^{12-14}$ low-frequency
constants also different.
We suggest that 
\par
\centerline{\psfig{figure=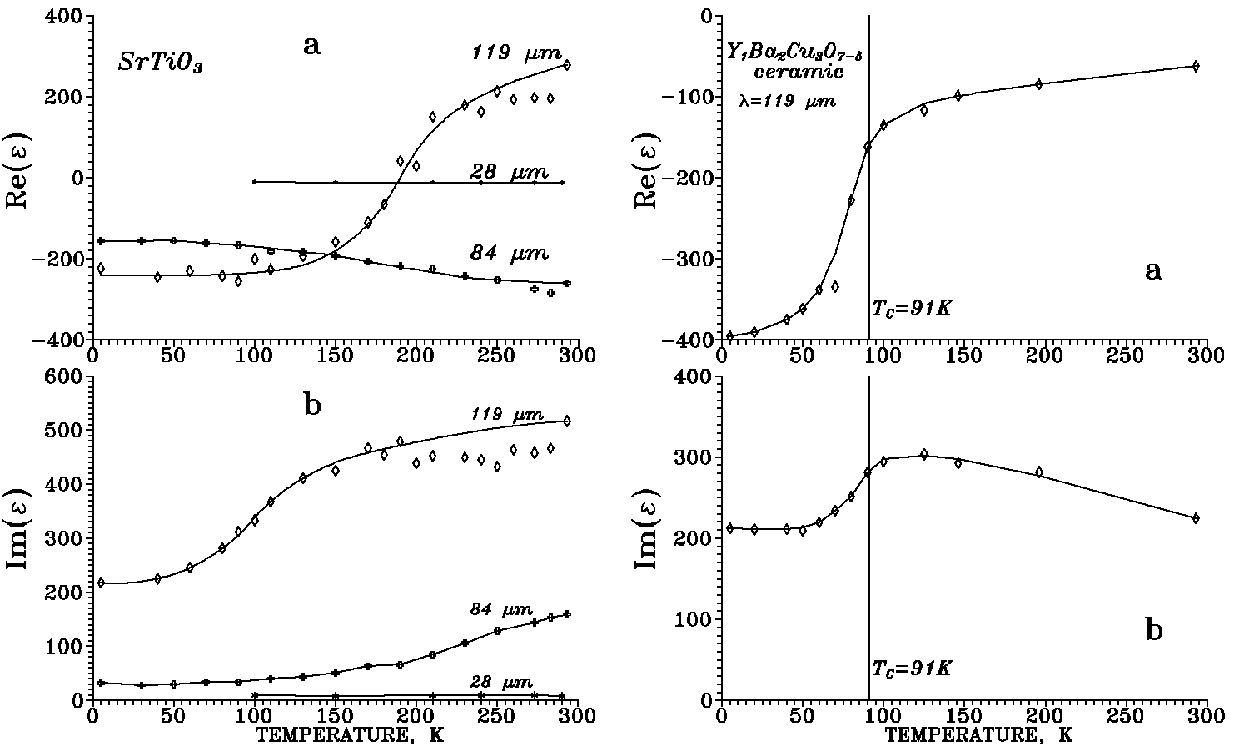}}
\small
\begin{minipage}[t]{60mm}
Fig. 5. Temperature dependence of the complex dielectric
function of SrTiO$_{3}$ single crystal at 3 wavelengths.
\end{minipage} \ \hspace{5mm} \
\begin{minipage}[t]{54mm}
Fig. 6. Temperature dependence of the complex dielectric
function of YBaCuO ceramic at wavelength 119 $\mu m$.
\vscone
\end{minipage} 
\normalsize
the lowest frequency mode is very sensitive to the
quality of SrTiO$_{3}$ single crystals.
\vscone \vscone \\
{\bf 4.3\ Superconductive transition in YBaCuO ceramic} \vscone 

Temperature dependence of the complex dielectric function of 1-2-3 ceramics
at wavelength 119 $\mu m$ is shown in Fig.~6. Sintered on air and annealed in oxygen
atmosphere sample was polished with diamond paste. The behavior of
the complex dielectric function is determined by chaotically oriented
small crystals with different optical constants for $a-$,
$b-$,~and~$c-$axis.
Porous structure of the surface of the ceramics also effects on the
measured values of the complex dielectric function.
\vsc
\newpage
\centerline{\bf 5.\ Appendix } \vscone

 Let 2$\times$2 matrix {\bf T} describes the transmittance of some window.
Rotations of this window around the axes give the following results:
\newlength{\aaa}
\settowidth{\aaa}{Rotation}
\begin{equation}
\begin{minipage}{\aaa}
     Rotation \\ \nolinebreak $x\rightarrow y$
\end{minipage}  \hspace{.5cm}
\left[ \begin{array}{rr}   0   &  1     \\  -1     &  0     \end{array} \right]
\left[ \begin{array}{cc} t_{11}& t_{12}  \\  t_{21}& t_{22} \end{array} \right]
\left[ \begin{array}{rr}   0   & -1      \\  1    &  0   \end{array} \right] =
\left[ \begin{array}{rr} t_{22}& -t_{21} \\ -t_{12}& t_{11} \end{array} \right]
\label{eq.S2-1}
\end{equation}
\begin{equation}
\begin{minipage}{\aaa}
     Rotation \\ \nolinebreak $y\rightarrow -y$
\end{minipage}  \hspace{.5cm}
\left[ \begin{array}{rr}   1   &  0     \\    0    & -1     \end{array} \right]
\left[ \begin{array}{cc} t_{11}& t_{12}  \\  t_{21}& t_{22} \end{array} \right]
\left[ \begin{array}{rr}   1   &  0      \\   0    & -1   \end{array} \right] =
\left[ \begin{array}{rr} t_{11}& -t_{12} \\ -t_{21}& t_{22} \end{array} \right]
\label{eq.S2-2}
\end{equation}
\begin{equation}
\begin{minipage}{\aaa}
     Rotation \\ \nolinebreak $x\rightarrow -x$
\end{minipage}  \hspace{.5cm}
\left[ \begin{array}{rr}  -1   &  0     \\   0     &  1     \end{array} \right]
\left[ \begin{array}{cc} t_{11}& t_{12}  \\  t_{21}& t_{22} \end{array} \right]
\left[ \begin{array}{rr}  -1   &  0      \\  0     &  1   \end{array} \right] =
\left[ \begin{array}{rr} t_{11}& -t_{12} \\ -t_{21}& t_{22} \end{array} \right]
\label{eq.S2-3}
\end{equation}
Transmission of the first two windows $W_{1}$ and $W_{2}$ can be written as:
\begin{eqnarray}
{\bf T}_{12}={\bf T}_{2} {\bf T}_{1}=
\left[ \begin{array}{cc} \tau_{11} & \tau_{12}  \\
                         \tau_{21} & \tau_{22}      \end{array} \right]
\left[ \begin{array}{cc}    t_{11} &    t_{12}  \\
                            t_{21} &    t_{22}      \end{array} \right] = & &
\nonumber \\
\left[ \begin{array}{cc}
\tau_{11}t_{11}+ \tau_{12}t_{21} & \tau_{11}t_{12}+ \tau_{12}t_{22}  \\
\tau_{21}t_{11}+ \tau_{22}t_{21} & \tau_{21}t_{12}+ \tau_{22}t_{22}
       \end{array} \right] & &
\label{eq.S2-4}
\end{eqnarray}
If windows $W_{1}$ and $W_{2}$ are identical and $W_{2}$ is rotated then
\begin{equation}
{\bf T}_{2}=
\left[ \begin{array}{rr} \tau_{11} & \tau_{12}  \\
                         \tau_{21} & \tau_{22}         \end{array} \right] =
\left[ \begin{array}{rr}    t_{22} & \pm t_{21}  \\
                        \pm t_{12} &     t_{11}         \end{array} \right]
\label{eq.S2-5}
\end{equation}
and, putting the off-diagonal elements much less then the diagonal ones,
\begin{eqnarray}
{\bf T}_{12}= \left[ \begin{array}{ll}
t_{11}t_{22} \pm t_{21}^{2}   & t_{22}(t_{12}\pm t_{21})  \\
t_{11}(t_{21}\pm t_{12})      & t_{11}t_{22} \pm t_{12}^{2}
       \end{array} \right] \approx  & &
\nonumber \\
    t_{11}t_{22} \left[ \begin{array}{ll}
                             1 & (t_{12}\pm t_{21})/t_{11}  \\
(t_{21}\pm t_{12})/t_{22}      & 1
       \end{array} \right], & &
\label{eq.S2-6}
\end{eqnarray}
where `$+$' means the presence and `$-$' means the absence of
$(x\rightarrow -x)$ or $(y\rightarrow -y)$ rotations.
So the values of the off-diagonal elements can be reduced to the minimum.
\vscone 

Let's now consider windows $W_{1}$ and $W_{2}$ as one window $W_{12}$ and
windows $W_{3}$ and $W_{4}$ as one window $W_{34}$ (Fig.~3B). Then
the reflection matrix ${\bf R}$ becomes:
\begin{eqnarray}
{\bf R}\rightarrow {\bf W}_{34} {\bf R}\, {\bf W}_{12}=
\left[ \begin{array}{cc} \tau_{11} & \tau_{12}  \\
                         \tau_{21} & \tau_{22}         \end{array} \right]
\left[ \begin{array}{cc}    R_{pp} &    R_{ps}  \\
                            R_{sp} &    R_{ss}         \end{array} \right]
\left[ \begin{array}{cc}    t_{11} &    t_{12}  \\
                            t_{21} &    t_{22}  \end{array} \right] \approx & &
\nonumber \\ 
\left[ \begin{array}{ll}
R_{pp}t_{11}\tau_{11} &
R_{ps}t_{22}\tau_{11}+R_{pp}t_{12}\tau_{11}+R_{ss}t_{22}\tau_{12}  \\
R_{sp}t_{11}\tau_{22}+R_{pp}t_{11}\tau_{21}+R_{ss}t_{21}\tau_{22}   &
R_{ss}t_{22}\tau_{22}    \end{array} \right] & &
\label{eq.S2-7}
\end{eqnarray}
Due to the rotations (\ref{eq.S2-1}),(\ref{eq.S2-2}) matrix ${\bf W_{34}}$
should be changed according to (\ref{eq.S2-5}) and the final expression for
the reflection matrix between two windows can be written as follows:
\begin{eqnarray}
{\bf W}_{34} {\bf R}\, {\bf W}_{12} = 
t_{11} t_{22} & &
\nonumber \\ 
\left[ \begin{array}{ll}
R_{pp} & (R_{ps}t_{22}+   R_{pp}t_{12}\pm R_{ss}t_{21}) / t_{11}  \\
         (R_{sp}t_{11}\pm R_{pp}t_{12}+   R_{ss}t_{21}) / t_{22}   &  R_{ss}
       \end{array} \right] & &
\label{eq.S2-8}
\end{eqnarray}
where, as usual, `$+$' means the presence and `$-$' means the absence of 
$(x\rightarrow -x)$ or $(y\rightarrow -y)$ rotations.
So, practically full compensation of the cryostat windows
effect can be reached. \vsc

\centerline{\bf 6.\ Acknowledgments} \vscone

 The authors thank 
G.~N.~Zhizhin and V.~A.~Yakovlev for the support
of this work and for the helpful discussions. \vsc

\centerline{\bf 7.\ References} \vscone
\makebox[3ex][r]{1.} \parbox[t]{11cm}{
R.~M.~Azzam and N.~M.~Bashara, ``Ellipsometry and
Polarized Light'', North-Holland, Amsterdam, 1977.} \ve

\makebox[3ex][r]{ 2.} \parbox[t]{11cm}{
A.~V.~Rjanov, K.~K.~Svitashev, A.~I.~Semenenko,
L.~V.~Semenenko, V.~K.~Sokolov, ``Osnovy ellipsometrii'', Novosibirsk, 1979.} \ve

\makebox[3ex][r]{ 3.}  \parbox[t]{11cm}{
R.~O.~DeNicola, M.~A.~Saifi, and R.~E.~Frazee,
``Epitaxial layer thickness measurement by far infrared ellipsometry'',
Appl. Opt., Vol. 11, No. 11, pp. 2534-2539, 1972.} \ve

\makebox[3ex][r]{ 4.}  \parbox[t]{11cm}{
A.~B.~Sushkov and E.~A.~Tishchenko,
``Ellipsometry of a convergent beam in the far infrared'',
Optika i Spectroscopia, Vol. 72, No. 2, pp. 491-496, 1992,
[Opt. Spectrosc. (USSR) Vol. 72, No. 2, pp. 265-268, 1992].} \ve

\makebox[3ex][r]{ 5.}  \parbox[t]{11cm}{
K.-L.~Barth and F.~Keilmann, ``Far-infrared ellipsometer'',
Rev. Sci. Instrum., Vol. 64, No. 4, pp. 870-875, 1993.} \ve

\makebox[3ex][r]{ 6.}  \parbox[t]{11cm}{
L.~A.~Vainshtein, ``Electromagnetic waves'', Moscow, 1988.} \ve

\makebox[3ex][r]{ 7.}  \parbox[t]{11cm}{
G.~Brandli and A.~J.~Sievers, ``Absolute measurement of the far-infrared
surface resistance of Pb'', Phys. Rev. B, Vol. 5, No. 11, pp. 3550-3557, 1972.} \ve

\makebox[3ex][r]{ 8.}  \parbox[t]{11cm}{
M.~A.~Ordal, L.~L.~Long, R.~J.~Bell, S.~E.~Bell,
R.~R.~Bell, R.~W.~Alexander,Jr., and C.~A.~Ward,
``Optical properties of metals Al, Co, Cu, Au, Fe, Pb, Ni, Pd, Pt,
Ag, Ti, and W in the infrared and far infrared '', Appl. Opt., Vol. 22, No. 7,
pp. 1099-1119, 1983.} \ve

\makebox[3ex][r]{ 9.}  \parbox[t]{11cm}{
J.~W.~Goodman, ``Introduction to Fourier Optics'',
McGrow-Hill, New-York, 1968.} \ve

\makebox[3ex][r]{10.}  \parbox[t]{11cm}{ D.~R.~Smith  and  E.~V.~Loewenstein,
``Optical constants of far infrared
materials. 3:plastics'', Appl. Opt., Vol. 14, No. 6, p. 1335, 1975.} \ve

\makebox[3ex][r]{11.}  \parbox[t]{11cm}{ V.~I.~Nijankovskii, Dr. Sci. Dissertation,
Institute for Physical problems, Moscow, 1990.} \ve

\makebox[3ex][r]{12.}  \parbox[t]{11cm}{
K.~F.~Pai, T.~J.~Parker, N.~E.~Tornberg, R.~P.~Lowndes,
and W.~G.~Chambers, ``Determination
of the complex refractive indices of solids in the far infrared by dispersive
Fourier transform spectroscopy \Roman{n}. Pseudo-displacive ferroelectrics'',
Infrared Phys., Vol. 18, No. 4, pp. 327-336, 1978.} \ve

\makebox[3ex][r]{13.}  \parbox[t]{11cm}{ J.~C.~Galzerani and R.~S.~Katiyar,
``The infrared reflectivity in SrTiO$_{3}$ and the antidistortive transition'',
Solid State Commun., Vol. 41, No. 7, pp. 515-519, 1982.} \ve

\makebox[3ex][r]{14.}  \parbox[t]{11cm}{
V.~N.~Denisov, B.~N.~Mavrin, V.~B.~Podobedov,
and J.~F.~Scott, ``Hyper-Raman spectra and frequency dependence of soft mode
damping in SrTiO$_{3}$'', J. Raman Spect. Vol. 14, No. 4, pp. 276-283, 1983.} \ve
\end{document}